\documentstyle[12pt,a4wide]{article}
\title{Cosmic Confusion}
\author{J.C.R. Magueijo \\ Department of Applied Mathematics and Theoretical
Physics\\University of Cambridge\\Cambridge CB3 9EW, UK }
\date{August 23, 1994}

\begin{document}
\bibliographystyle{unsrt}

\maketitle
\begin{abstract}
We propose to minimise the cosmic confusion between Gaussian and non Gaussian
theories by investigating the structure in the m's for each multipole of the
cosmic radiation temperature anisotropies. We prove that Gaussian
theories are (nearly) the only theories which treat all the m's equally.
Hence we introduce a set of invariant measures of ``m-preference''
to be seen as non-Gaussianity indicators.
We then derive the distribution function for the quadrupole ``m-preference''
measure in Gaussian theories. A class of physically motivated toy
non Gaussian theories is introduced as an example.
We show how the quadrupole
m-structure is crucial in reducing the confusion
between these theories and Gaussian theories.
\end {abstract}
\vspace{8cm}
\begin{flushright}
Revised Version
\end{flushright}
\thispagestyle{empty}
\pagebreak

\pagestyle{plain}
\setcounter {page}{1}
\section{Introduction}
Several competing theories of the Early Universe have now been
proposed. Overall, two major families have emerged, one centred around
the idea of cosmic inflation $\cite{kbt}$, the other relying on topological
defects formed in phase transitions $\cite{txpen}$
$\cite{vil}$. Crudely speaking, it was found
that the first gives rise to a Universe with Gaussian features,
whereas the latter should leave some sort of non-Gaussian imprint
in the Universe. Producing a cosmic crucible has however been problematic.
The discovery of temperature fluctuations in the cosmic microwave
background (CMBR) opened doors to a new field with a
chance for a decisive
experiment. In this letter we examine the obstacles cosmic variance
raises to the design of such an experiment $\cite{cvar1}$
$\cite{cvar2}$, and we propose
a new approach to circumventing them.

Cosmic variance in the CMBR comes about because for the Universes
predicted by every theory (and probably for our Universe too)
there is only a probability distribution function for what the CMBR sky
should look like. Different observers, located at different,
uncorrelated, points
of the Universe, observe different skies. These follow
a distribution function which varies from theory to theory,
but which is never so peaked that only one sky is possible.
The distributions naturally overlap, and so cosmic confusion arises.
Mere mortals, with access to only one sky, may be left with the
problem that their sky could have derived from totally different theories.
To be quantitative, let us define cosmic confusion as the percentage
of common skies generated by two theories. Then, if a measurable quantity
$Q$ is predicted to have distributions $F_1(Q)$ and $F_2(Q)$ in the two
theories $T_1$ and $T_2$, the cosmic confusion in $Q$ between $T_1$ and
$T_2$ is
\begin{equation}
{\cal C}_Q(T_1,T_2)=\int_{all\; Q}dQ\; \min (F_1(Q),F_2(Q))\; .
\end{equation}
${\cal C}_Q(T_1,T_2)$ varies between $0$ (no doubts) and
$1$ (totally confused), and is the confusion before $Q$ is actually
measured. It is the quantity to be minimised (by means of
an appropriate choice of $Q$) when projecting experiments which,
one hopes, will be conclusive. Once $Q$ is measured, say,
with an outcome
$Q\in {\cal D}_Q$, ${\cal D}_Q=(Q_0-\Delta Q^-,Q_0+\Delta Q^+)$,
the confusion becomes
\begin{equation}
{\tilde{\cal C}}_Q(T_1,T_2)={2{\int_{{\cal D}_Q}dQ\; \min (F_1,F_2)}
\over {\int_{{\cal D}_Q}dQ\; (F_1+F_2)} }
\end{equation}
and if $ {\int_{{\cal D}_Q}dQ\; F_1} >{\int_{{\cal D}_Q}dQ\; F_2}$,
the probability of $T_1$ over $T_2$ is
\begin{equation}
P_Q (T_1,T_2) ={{\int_{{\cal D}_Q}dQ\; F_1}\over
{\int_{{\cal D}_Q}dQ\; (F_1+F_2)} }\; .
\end{equation}
The attitude adopted in this letter is alarmingly anti-anarchist.
We seek to minimise the cosmic confusion between Gaussian and non Gaussian
theories using quantities associated with the low order multipoles of
${\delta T\over T}$ (as opposed to $l>30$, where confusion is known to
be small again). Rather than looking at the angular power spectrum
$C^l$ or the quadrupole intensity $Q_{rms}$, we investigate the
structure in the m's for each $l$ (we will actually concentrate
on $l=2$ in this letter). Gaussian theories are known to treat all m's
in the same way. In Section $\ref{theorem}$
we prove that such a feature
is in fact peculiar to Gaussian theories,
within a large class of theories.
The physically most relevant theories, however, appear to be
outside the class considered, but in Section $\ref{crash}$
we show how ``m-preference''  still plays an important role
as non-Gaussianity indicator in the general case.
Hence quantities measuring
m-preference appear as good candidates in our quest for low cosmic
confusion. In Section $\ref{mtpole}$
we write down the most general
multipole invariants and identify a set of invariant measures
of m-preference. We then specialise to the quadrupole
(Section $\ref{quadinv}$) and
in Section $\ref{Gaussquad}$
the distribution function for its measure of m-structure is derived
for Gaussian theories.
A class of physically motivated toy non Gaussian theories is
introduced in Section $\ref{toynonGauss}$.
We exhibit one extreme case where the confusion in
$Q_{rms}$ is 1, but where the confusion in the quadrupole
m-structure is $0$.
We conclude with a brief summary of the
results obtained.

\section{Structure in the m's as a sign of non Gaussianity}\label{theorem}

In analyzing ${\delta T\over T}(\theta,\phi)$ maps it is traditional
to use the multipole expansion:
\begin{equation}
{\delta T\over T}=\sum_{l=0}^{\infty} \sum_{m=-l}^l a^l_m Y^l_m =
\sum_{l=0}^{\infty} \sum_{m=-l}^l b^l_m Q^l_m \; .
\end{equation}
The real and complex spherical harmonics ($Q^l_m$ and $Y^l_m$) are
related by
\begin{equation}\label{rel1}
Y^0_0=Q^0_0\quad ;\quad Y^l_{\pm m}=(\mp)^m{Q^l_m\pm iQ^l_{-m}\over \sqrt 2}
\end{equation}
and similar relations apply to their coefficients $b^l_m$ and $a^l_m$.
In this Section we assume that the $b^l_m$ are independent
random variables.
Under this condition we prove a rather surprising result: that only Gaussian
theories
are statistically spherically symmetric (SSS) and m-amorphous
(i.e: treat all the $b^l_m$ in the same way for each $l$).
The implication is that conversely any non-Gaussian theory is to some
extent anisotropic, favouring particular directions in the sky
and some m's over the others.

We say that a theory is SSS if any two quantities in the sky
related by an
$O(3)$ transformation are predicted to have the same distribution
function ($f(R(Q))=f(Q)$). We also say that a theory is m-amorphous
if for a given fixed $l$ the $b^l_m$ are independent random variables
with the same distribution function, regardless of $m$
and the choice of axes.
If this distribution is Gaussian we then say that the theory is
Gaussian. Let us first prove the following Lemma:
\newtheorem{lemma}{Lemma}
\begin{lemma}
Let $x$ and $y$ be two independent continuous random variables with
the same distribution $f$.
Let $z=\alpha x+\beta y$ with $\alpha^2 +\beta^2=1$.
Then, if $f$ is also the $z$
 distribution,
$f$ must be a Gaussian.
\end{lemma}
${\bf Proof:}$
Let $\phi$ be the characteristic function of $f$
(e.g. $\cite{stat1}$). Then
\begin{equation}
\phi(t)=\phi(\alpha t)\phi(\beta t)\; .
\end{equation}
Putting $g=\log f$ this implies that
\begin{equation}\label{c1}
g(t)=g(\alpha t) + g(\beta t)\; .
\end{equation}
Now expand $g$ in power series:
\begin{equation}
g(t)=\sum g_n t^n\; .
\end{equation}
Then ($\ref{c1}$) requires that
\begin{equation}
g_n=(\alpha^n +\beta^n)g_n
\end{equation}
and so $g_n\neq 0$ iff
\begin{equation}\label{cvs}
\alpha ^n +\beta^n =1\; .
\end{equation}
But we know that $\alpha^2 +\beta^2=1$, and by studying the curves
($\ref{cvs}$) in the $(\alpha,\beta)$ plane for all the $n$'s one finds
that they only intersect at $\alpha=0$ or $\beta=0$. Hence $n=2$, and so
$g(t)\propto t^2$ and $\phi\propto e^{g_2t^2}$, which is the
characteristic
of a Gaussian distribution.(Q.E.D.)

We can now prove the central result of this Section.
\newtheorem{theorem}{Theorem}
\begin{theorem}
For a theory in which all the $b^l_m$ are continuous independent
random variables the following three properties are equivalent: SSS,
m-amorphism, and Gaussianity.
\end{theorem}

{\bf Proof:} It suffices to show that SSS implies Gaussianity.
(It looks as if we are proving more than the equivalence of
m-amorphism and Gaussianity because it is not immediately obvious that
a SSS theory is also m-amorphous.
This is because given a basis $\{ Q^l_m\}$ its elements
cannot all be $O(3)$-transformed into each other.)
We start by noting that by rotating an angle $\phi$ around $z$, for a given
fixed $m\neq 0$, the elements $Q^l_m$ and $Q^l_{-m}$ transform like
\begin{eqnarray}
Q^l_m&\rightarrow&{\tilde Q}^l_m=\alpha Q^l_m -\beta Q^l_{-m}\nonumber\\
Q^l_{-m}&\rightarrow&{\tilde Q}^l_{-m}=\beta Q^l_m +\alpha Q^l_{-m}
\end{eqnarray}
with $\alpha=\cos\phi$, and $\beta=\sin\phi$. Choosing $\phi=\pi/2$
we then know that $f(b^l_m)=f(b^l_{-m})=f$.
Since also $f({\tilde b}^l_m)=f$ for any $\phi$,
using the Lemma we know that $f$
has to be a Gaussian. Now, let us apply the most general rotation to
$Q^l_m$. Any element of its orbit will have the distribution function
$f$. The subspace spanned by the orbit is not the whole space, but
because the representation is irreducible, it will not be an
invariant subspace. This means that vectors $V$ outside the subspace
exist
which, although not obtainable directly by a rotation of $Q^l_m$,
can still be obtained by rotating a linear combination of rotations
of $Q^l_m$:
\begin{equation}
V=R[\sum_n\alpha_n R_n(Q^l_m)]\; .
\end{equation}
By choosing $\sum\alpha^2_n=1$ one finds that $f(V)=f$.
$V$ and the orbit of $Q^l_m$ still do not span
the whole space, but the procedure described has allowed us
to extend non trivially the subspace spanned by the elements
for which the distribution is
$f$. By iteration one can therefore show that for all $m$
the distribution $f(b^l_m)$ is a Gaussian
distribution function $f$ with a variance which can only
depend on $l$.(Q.E.D.)

\section{SSS non-Gaussian theories}\label{crash}

The theories for which all the $b^l_m$ are independent
form a large class. Within this class m-structure is a
certain indicator of non-Gaussianity. However only non-SSS
non-Gaussian theories can belong to this class. While violations
of SSS have appeared in proposed models (such as the $\delta
T\over T$ brought about by any anisotropic cosmological model),
one should beware of these violations as they
can never be tested experimentally. Furthermore most practical
applications concern $\delta T\over T$ due to perturbations in
FRW models, for which SSS is always satisfied. If one takes SSS as
a starting point, then one must confront the following corollary
of Theorem 1.
\newtheorem{cor}{Corollary}
\begin{cor}
For SSS non-Gaussian theories the $b^l_m$ can never be all independent
random variables.
\end{cor}
Once one allows the $b^l_m$ to be dependent
the rest of Theorem 1 breaks down
(see the end of Section $\ref{Gaussquad}$ for
an example). For SSS theories
m-structure may or may not be associated
with non-Gaussianity. In some cases m-structure is still the hallmark
of non-Gaussianity (see Section $\ref{toynonGauss}$), but there is also
a class of non-Gaussian m-amorphous theories.

However we now show that if a non-Gaussian theory is m-amorphous
and sufficiently distant in functional space from Gaussian theories,
then the $C^l$ must be distinctly non-Gaussian. The power spectrum
appears to save the day when m-structure is useless. We show
this fact with the aid of the concept of usefulness of a variable.
Let the distance between two theories be
\begin{equation}
{\cal D}(T_1,T_2)=\int_{all\; variables} |F_1 - F_2|\in (0,2)\; .
\end{equation}
For any subset of variables ${\cal S}$ we have
\begin{equation}
1-{\cal D}/2\le{\cal C}_{\cal S} (T_1,T_2)\le 1\; ,
\end{equation}
showing that the minimal possible confusion is
${\cal C}_{min}= 1-{\cal D}/2$. If ${\cal D}=0$,
the two theories are the same, and no wonder we are totally confused
whatever set of variables we look at. For ${\cal D}\neq 0$,
however, one can assess the usefulness of the variables ${\cal S}$
with the quantity
\begin{equation}
{\cal U}_{\cal S}=1-{{\cal C}_{\cal S}-{\cal C}_{min}\over
1-{\cal C}_{min}}\; .
\end{equation}
${\cal U}_{\cal S}$ varies between 0 (utterly useless) to 1 (perfect
choice). Obviously if ${\cal S}$ contains all the variables,
${\cal U}_{\cal S}=1$, but what we should be after is maximising
${\cal U}$ with the smallest number of variables.
Now if $T_1$ and $T_2$ are two m-amorphous theories (Gaussian or not)
we have $F(b^l_m)=f(C^l)$ and so
\begin{equation}
{\cal D}=\int (\prod_{lm} db^l_m) |F_1(b^l_m)-F_2(b^l_m)|
=\int (\prod_l dC^l)|F_1(C^l)-F_2(C^l)|\; .
\end{equation}
Therefore ${\cal U}_{C^l}=1$, implying that if $T_1$ is Gaussian
and ${\cal D}(T_1,T_2)\approx 2$ then ${\cal C}_{C^l}(T_1,T_2)
\approx 0$, that is, $C^l$ for $T_2$ must be distinctly non-Gaussian.

The important converse statement of the last paragraph
is that a very non-Gaussian theory with a very Gaussian
$C^l$ must display a distinctly non-Gaussian m-structure. This
clarifies the role of m-structure for SSS theories:
m-structure may or may not work for them as a non-Gaussianity
indicator, but it always works when $C^l$ does not.

\section{Multipole invariants and structure in the m's}
\label{mtpole}

These results prompt us to quantify m-preference as
m-preference measures will also measure non-Gaussianity.
For this purpose it will be useful to first recall
the isomorphism between the space spanned by
the real spherical harmonics of degree $l$ and the space of the real
traceless symmetric cartesian tensors of rank $l$
($Q_{i...j}=Q_{(i ...j)}$, $Q_{iij...k}=0$).
This isomorphism can best be
established by writing the functions $Q^l_m(\theta,\phi)$ in cartesian
coordinates. The resulting polynomial of degree $l$ has the form
$Q^{lm}_{i...j}x^i...x^j$, identifying
a set of traceless symmetric tensors
of rank $l$:
\begin{equation}\label{ismp}
\{ Q^l_m (\theta,\phi)\}\leftrightarrow\{ Q^{lm}_{i...j}\}\; .
\end{equation}
The isomorphism can then be extended to the
rest of the space using linearity. The basis $\{Q^{lm}_{i...j}\}$
is known as an irreducible tensor set and one can check that
indeed both spaces are $2l+1$ dimensional
(see for instance $\cite{ham}$).

Now for whatever combination of the $a^l_m$ (or $b^l_m$) one
considers it is important to demand spherical invariance,
so as to factor out the artifacts introduced
by our inevitable choice of axes.
Hence we look for multipole invariants.
Under an axes transformation
defined by the Euler angles $(\psi,\theta,\phi)$ the $a^l_m$
transform according to
\begin{equation}
a^l_m\rightarrow {\tilde a}^l_m =\sum_{m'}D^l_{m'm}a^l_{m'}
\end{equation}
where $D^l_{m'm}(\psi,\theta,\phi)$ is the Wigner matrix.
The tensors $Q_{i...j}$, on the other hand, transform just
like any other
cartesian tensor. The fact that the matrices $D^l_{m'm}$ form
an irreducible representation of the 3-dimensional rotation
group requires, by means of Schur's Lemma, that the only
matrices they commute with be multiples of the identity.
As a result, the only independent
${\em bilinear}$ invariant one can construct out
of the $a^l_m$ (or $b^l_m$) is
\begin{equation}
C^l=\sum_{m=-l}^l {\left| a^l_m\right|}^2=\sum_{m=-l}^l(b^l_m)^2
\end{equation}
more proverbially known as the angular power spectrum.
However, by simply counting degrees of freedom, one is easily
convinced that there are in fact $2l-2$
invariants for each $l\geq 2$. What Schur's Lemma states is that
$2l-3$ of them will not be bilinear (that is, they will
be of higher order). Therefore, the invariants are in general
multilinear forms
\begin{equation}
I^{(n)}=\sum_{i...j=1}^{2l+1} I^{(n)}_{i...j} a^l_i...a^l_j
\end{equation}
satisfying the condition
\begin{equation}
I^{(n)}_{i'...j'}=D^l_{i'i}...D^l_{j'j} I^{(n)}_{i...j}\; .
\end{equation}
These are very difficult to find for a general $l$. Since
the rotation matrices $D^l_{m'm}$ are also a subgroup of
$SO(2l+1)$ we know that $l$ of these invariants will be the
fundamental
representations of the Casimir operators of $SO(2l+1)$.
Still, even these have a rather unpleasant form.
We have found it easier to write the invariants in terms
of the tensors $\{ Q_{i...j}\}$ instead. They then appear
as the independent contractions of a symmetric traceless tensor
of rank $l$ in 3 dimensions. We cannot produce
a general formula for these, but
this approach has proved to be more manageable
case by case.

What is the meaning of the $I^{(n)}$? The invariant $C^l$
is obviously a measure of the overall intensity of the multipole.
It treats all m's equally, so it does not assess m-preference at all.
The other $I^{(n)}$, however, treat the various m's
differently (see Section $\ref{quadinv}$ for an example). For all invariants
of order $m>2$ one should then define the ratios:
\begin{equation}
r^{(n)}={I^{(n)}\over (C^l)^{m/2}}\; .
\end{equation}
The shape factors $r^{(n)}$ reveal how some m's are
preferred over others, or, in other words, how much and
what type of anisotropy there is in the multipole, but
only in so far as there is a rotationally invariant meaning to the
concept. Demanding rotational invariance is important since
otherwise our anisotropy measures would reflect not only
the directionality pertaining to the CMBR but also the directionality
imparted by our choice of axes. To make the point clear
consider the case $l=1$. Using ($\ref{ismp}$)
the dipole can be seen as a vector,
for which the only invariant is $C^1$ (the dipole has no shape).
Clearly, $any$ dipole has a direction, so it could never distinguish
between an isotropic and an anisotropic theory.

There are also invariants which combine different $l$'s.
By counting degrees of freedom we find that, for each pair
$l,l'\geq 2$, there are 3 inter-$l$ invariants which do not depend
on the invariants for each $l$. We may wish to look at $I^{(n)}$
as generalised eigenvalues. For a general $l$ one can define the
eigenvectors of $Q_{i...j}$ as the set of axes where three independent
prechosen components of $Q_{i...j}$ are set to zero (for $l=2$,
$Q_{ij}=0$ for $i\neq j$), and its eigenvalues as the values of
the remaining components in that basis (for $l=2$ the diagonal components).
The inter-$l$ invariants can then be interpreted as the Euler
angles of one set of eigenvectors with respect to the other.
They are uniformly distributed in Gaussian theories,
but not in defect theories, where they reveal the correlations
between successive generations of defects $\cite{ltx}$.

\section{Quadrupole variables and invariants}\label{quadinv}

We now specialise to $l=2$.
Using the isomorphism ($\ref{ismp}$) a quadrupole can be seen as a
traceless symmetric real matrix $Q_{ij}$. This can be fully characterised
by the Euler angles $(\psi ,\theta ,\phi)$ specifying the
rotation required to diagonalize $Q_{ij}$, together with
its diagonal form
${\hat Q}_{ij}$. The matrix $Q_{ij}$ has two invariants under
$O(3)$ transformations, which can be parameterized in a variety of ways.
The eigenvalues
${\hat Q}_{ij} = \;{\rm diag} (\lambda_1,\lambda_2,\lambda_3)$
(subject to $\lambda_1+\lambda_2+\lambda_3=0$),
if considered in modulus and in unordered triplets, constitute one such
parameterization (recall that improper rotations permutate the eigenvalues).
One can also expand  ${\hat Q}_{ij}$
in multipoles:
\begin{equation}
{\hat Q}_{ij}=a{\hat Q}_{ij}^0 +b{\hat Q}_{ij}^2\; .
\end{equation}
Again, $a$ and $b$ are not left unchanged by improper rotations,
but transform according to
\begin{eqnarray}\label{syr}
(1)&&a\rightarrow -a\nonumber\\
(2)&&b\rightarrow -b\nonumber\\
(3)&&
a\rightarrow {{\sqrt 3}b-a\over 2}\nonumber\\
&&b\rightarrow{{\sqrt 3}a+b\over 2} \; .
\end {eqnarray}
Modulo these transformations, $a$ and $b$ also
parameterize the
$Q_{ij}$ invariants. Finally, one can
do away with any residual transformations
by considering the invariants:
\begin{eqnarray}
I_1=\;{\rm Tr} Q^2=Q_{ij}Q^{ij}=\sum \lambda^2_i&=&
{30\over 16\pi}\sum b_i^2
={30\over 16\pi}(a^2+b^2)\\
I_2={\left| {1\over 3}\;{\rm Tr} Q^3\right|}={\left|\det Q\right|}
=\prod\lambda_i&=&{\left(15\over 16\pi\right)}^{3/2}
\left|\begin{array}{ccc}
{2b_0\over {\sqrt 3}}&b_1&b_{-1}\\
b_1&-{b_0\over {\sqrt 3}}+b_2&b_{-2}\\
b_{-1}&b_{-2}&-{b_0\over {\sqrt 3}}-b_2
\end {array}\right| =\nonumber \\
&=&{\left(15\over 16\pi\right)}^{3/2}
{\left|{2a\over{\sqrt 3}}{\left( {a^2\over 3} -b^2\right)}\right|}\; .
\end{eqnarray}
We should note that since in a SSS theory $\{\lambda_i\}$
and $(a,b)$ are subjected to
residual symmetries interconnecting them they can
never be independent
random variables. In fact their joint distribution functions,
subjected to the same symmetries, can never factorize
(cf. ($\ref{abdist}$)). Therefore,
for statistical applications, they are not good
parameterizations of the quadrupole invariants. Also,
one can prove that
\begin{equation}
{|\lambda_1\lambda_2\lambda_3 |\over {\left(\lambda_1^2
+\lambda_2^2 +\lambda_3^2\right)}^{3/2}} \leq {1\over 3{\sqrt 6}}
\end{equation}
and so we have the constraint
\begin{equation}
I_2\leq{I_1^{3/2}\over 3{\sqrt 6}}\; .
\end{equation}
Consequently, even though $I_1$ and $I_2$ are not
interconnected by residual
symmetries (and so their joint distribution function does
factorize), their ranges of variation are
dependent ($I_1\in (0,\infty)$,
$I_2\in(0,{I_1^{3/2}\over 3{\sqrt 6}})$). Hence $I_1$ and
$I_2$ are in fact
dependent random variables, a fact
manifest in that
the integrated distribution function of $I_2$, ${\tilde F}(I_2)=
\int F(I_1,I_2)dI_1$,
is different from the factor $F_2(I_2)$ appearing in
$F(I_1,I_2)=F_1(I_1)F_2(I_2)$. For this reason we shall characterize
the quadrupole $Q_{ij}$ by the Euler angles $(\psi,\theta,\phi)$,
the intensity
$I_1$ (known as $Q_{rms}$), and the ratio $r$
(to be called the ``quadrupole shape''):
\begin{equation}
r=3{\sqrt 6}{I_2\over I_1^{3/2}}
\end{equation}
which varies in the range $r\in(0,1)$. These five
quantities are statistically
independent: their joint distribution function factorizes
and their ranges
of variation are independent. The shape factor $r$ is an invariant measure
of how axis-symmetric the quadrupole is. It varies from
maximal symmetry for $r=1$ (two equal $\lambda$; $b=0$ possible)
to maximal symmetry breaking for $r=0$ (a null $\lambda$; $a=0$
possible).

\section{The distribution of the quadrupole shape in Gaussian
theories}\label{Gaussquad}

It turns out to be easier to find first the distribution
function $F(a,b,\psi,\theta,\phi)$. Since
Gaussian theories are SSS we know
that $F$ has to have the form
\begin{equation}\label{form}
F(a,b,\psi,\theta,\phi)=F(a,b){\sin\theta\over 8\pi^2}
\end{equation}
where $F(a,b)$ is further invariant under the transformations ($\ref{syr}$).
Then start by writing:
\begin{eqnarray}\label{det}
F(a,b,\psi,\theta,\phi)&=&F(b_0,b_1,b_{-1},b_2,b_{-2}){\left|
{\partial (b_0,b_1,b_{-1},b_2,b_{-2})\over\partial
(a,b,\psi,\theta,\phi)}\right|}=\nonumber \\
&=&{{\exp {\left(-{a^2 +b^2\over 2\sigma_2^2}
\right)}}
\over (2\pi)^{5/2} \sigma_2^5}.{\left|
{\partial (a_0,a_1,a_{-1},a_2,a_{-2})\over\partial
(a,b,\psi,\theta,\phi)}\right|}\; .
\end{eqnarray}
Now let $D^2_{mm'}(\psi,\theta,\phi)$ be the
Wigner matrix of the rotation
taking ${\hat Q}_{ij}$ to $Q_{ij}$. Then
$a^2_m=aD^2_{m0}+bH^{2+}_{m}$ where
$H^{2\pm}_m=(D^2_{m2}\pm D^2_{m-2})/ \sqrt 2$.
The matrices $D^2_{mm'}$ can be seen as generalised spherical
harmonics satisfying differential equations similar to the
spherical harmonics equations $\cite{gel}$. In particular:
\begin{eqnarray}
{\partial D^2_{m0}\over\partial\phi}&=&-imD^2_{m0}\nonumber\\
{\partial D^2_{m0}\over\partial\psi}&=&0\nonumber\\
{\partial D^2_{m0}\over\partial\theta}&=&m\cot\theta D^2_{m0}
+{\sqrt{ (2-m+1)(2+m)}} . e^{i\phi} . D^2_{m-1\; 0}
\end {eqnarray}
and
\begin{eqnarray}
{\partial H^{2\pm}_m\over\partial\phi}&=&-imH^{2\pm}_m\nonumber\\
{\partial H^{2\pm}_m\over\partial\psi}&=&-2iH^{2\mp}_m\nonumber\\
{\partial H^{2+}_m\over\partial\theta}&=&2{\cos\theta} H^{2+}_m
-{m\over \sin\theta}H^{2-}_m \; .
\end {eqnarray}
These relations imply that the determinant in ($\ref{det}$)
must have the form
\begin{equation}
{\left|{\partial (a_0,a_1,a_{-1},a_2,a_{-2})\over\partial
(a,b,\psi,\theta,\phi)}\right|}=a^2b\phi_1(\psi,\theta,\phi)
+ab^2\phi_2(\psi,\theta\,\phi) +b^3\phi_3(\psi,\theta\,\phi)
\end{equation}
and its exact form can then be found
from the symmetries.
In fact ($\ref{form}$) implies that
$\phi_1,\phi_2,\phi_3\propto\sin\theta$
and  ($\ref{syr}$) leads to:
\begin{equation}\label{abdist}
F(a,b,\psi,\theta,\phi)= C{\sin\theta\over 8\pi^2}
{\exp {\left(-{a^2 +b^2\over 2\sigma_2^2}\right)}}
{\left|b(3 a^2 - b^2)\right|}\; .
\end{equation}
Having found $F(a,b,\psi,\theta,\phi)$ one can now write
$F(I_1,I_2,\psi,\theta,\phi)$ as
\begin{equation}
F(I_1,I_2,\psi,\theta,\phi)=\sum_{8\; branches}F(a,b,\psi,\theta,\phi)
{\left| \partial(a,b)\over\partial (I_1,I_2)\right|}\; .
\end{equation}
The branches referred to under the summation sign
are the 8 branches generated by the symmetries ($\ref{syr}$)
which give the same values ($I_1,I_2$). It can be proved
that not only $F(a,b,\psi,\theta,\phi)$ but also the determinant
${\left| \partial(a,b)\over\partial (I_1,I_2)\right|}$
are invariant under ($\ref{syr}$). Hence all 8
branches give the same result
which can be more easily evaluated for the $a>{\sqrt 3} b>0$
branch. One then gets
\begin{equation}
F(I_1,I_2,\psi,\theta,\phi)=C^{'}{\sin\theta\over 8\pi^2}.
e^{-{I_1\over
2\tilde\sigma_2^2}}
\end{equation}
with ${\tilde \sigma_2}=\sigma_2{\sqrt{30\over 16\pi}}$, and from it
one finally obtains
\begin{equation}\label{fdist}
F(I_1,r,\psi,\theta,\phi)={e^{-{I_1\over
2\tilde\sigma_2^2}} I_1^{3/2}
\over 3{\sqrt{2\pi}}{\tilde\sigma_2^3}}
.{\sin\theta\over 8\pi^2}
\end{equation}
where the proportionality constant was computed from a normalization
condition. We have thus determined the distribution of the 5
independent variables of the quadrupole in a Gaussian theory.
The quadrupole intensity
$I_1$ is a $\chi^2_5$ variable (as well known), its shape $r$ is uniformly
distributed in $(0,1)$ and its axes variables
$(\psi,\theta,\phi)$ have a uniform
distribution. We have confirmed these results with a Monte-Carlo simulation.

We can now, as a side remark, provide the example
required in Section $\ref{crash}$. Note that for $l=2$, SSS
(equivalent to $F(I_1,r,\psi,\theta,\phi)=$ $f(I_1,r)
{\sin\theta\over 8\pi^2}$) implies that $F(b_0,b_1,b_{-1},b_2,b_{-2})$
$\propto $ $ f(I_1,r)/I_1^{3/2}$. On the other hand,  m-amorphism
(equivalent to $F(b_0,b_1,b_{-1},b_2,b_{-2})=$ $f(I_1)$) implies
that $F(I_1,r,\psi,\theta,\phi)\propto$ $ f(I_1)I_1^{3/2}\sin\theta$.
These general formulae show that, unless we require that
$F(b_0,b_1,b_{-1},b_2,b_{-2})$ factorizes, SSS does not imply
m-amorphism, and neither implies Gaussianity.

\section{A (toy) non-Gaussian theory}\label{toynonGauss}
We now produce a toy non-Gaussian theory in order to show
how m-structure measures can minimize
the confusion between
Gaussian and non-Gaussian theories. Imagine a theory which imprints
a quadrupole in ${\delta T\over T}$ such that there is always
an eigenvector basis where
\begin{equation}\label{ptquad}
{\tilde Q}_{ij}=aQ^{(20)}_{ij}\; ,
\end{equation}
but the eigenvectors' orientation is uniformly distributed.
This toy model can be physically motivated.
It is a very good approximation
to the quadrupole brought about  by a B-field with
a very large coherence length (see $\cite{sol}$ or $\cite{mag}$).
It is also a good approximation to the
quadrupole predicted in a texture scenario
with scaling and with a small average number
of defects per horizon volume ($<N>\, \ll 1$). Reasoning in accordance with
$\cite{ld1}$ one should expect the quadrupole in this scenario to result
mostly from the gravitational
effects of $the$ last defect.
If most of the collapses are reasonably spherical the quadrupole
will then be approximated by ($\ref{ptquad}$).

The quadrupole invariants for this
theory are
\begin{eqnarray}
I_1&=&{30\over 16\pi} a^2\nonumber\\
r&=&1\; .
\end {eqnarray}
Consider then the extreme situation
in which $a$ is such that $I_1$ is $\chi^2_5$-distributed.
The other quadrupole variables follow the distributions
\begin{eqnarray}
f(r)&=&2\delta (r-1)\nonumber\\
f(\psi,\theta,\phi)&=&{\sin\theta\over 8\pi^2}\; .
\end {eqnarray}
Hence, the cosmic confusion in the quadrupole invariants between
this theory ($T_2$) and a Gaussian theory ($T_1$) with the same $I_1$
is
\begin{eqnarray}
{\cal C}_{I_1}(T_1,T_2)&=&1\nonumber \\
{\cal C}_r(T_1,T_2)&=&0\; .
\end {eqnarray}
The distributions $f_{T_2}(b_n)$ are somewhat complicated (they will
be derived and used in $\cite{ltx}$), but it should be obvious that they do
not reduce the confusion to $0$. The invariant $r$ is clearly the
appropriate variable to thoroughly make out the difference between the
two theories. Using the concept of usefulness ${\cal U}_{C^l}=0$
and ${\cal U}_r=1$.

Naturally one should not expect to find such an extreme situation, say,
in the texture scenario. First, the confusion in $I_1$ is
probably not as big, and second $r$ will peak around $1$ but
not so drastically. The current all-sky simulations have not yet
produced a sufficient number of skies for distribution functions
to be predicted with any statistical relevance (only 8 skies
were produced in $\cite{txpen}$). In $\cite{ltx}$
we shall describe a computational
shortcut which will allow a prediction for
$F_{tex}(I_1)$ and $F_{tex}(r)$.

\section{Summary and planned developments}
In this letter we proposed yet another non-Gaussianity indicator:
m-structure (see $\cite{luo}$ for a summary of other options).
We gave general arguments linking m-structure and
non-Gaussianity, and set up the general framework for quantifying
m-structure. We wrote down explicit measures for the quadrupole
m-structure, and derived their distribution in Gaussian theories.
We finally displayed physically
motivated examples of theories for which only m-structure can
bring out non-Gaussian nature.

The signal to noise ratio in current all-sky data is not yet
good enough to allow any meaningful measurement of $r$
(still we hope to apply our approach to COBE data analysis).
Nevertheless, an accurate measurement of $r$ meets no obstacles
of principle, with the possible exclusion of the effects of
galactic obscuration. It is not clear how much of a problem this
is, but we hope to generalise our work to $l>2$,
so as to minimise this effect.

In work currently in progress we also seek to derive an approximation
to $f(r)$ in various defect theories. Preliminary results seem to
generally link the quadrupole with one individual defect (the closest
defect). That being the case, the defect morphology has an imprint
on the quadrupole shape, as the defect symmetries impose selection
rules on the spherical harmonics expansion. A rich peak structure
in $f(r)$, depending on the defect type, ensues. This supports the
claim that $r$, if measurable, is a low confusion variable
when confronting defect theories among themselves and against
Gaussian theories.

\vspace{.1 cm}
{\bf ACKNOWLEDGMENTS } I would like to thank N.Turok, K.Baskerville
and the referee
for helpful comments.
I should also thank everyone
at the Department of
Physics of Princeton University, where this work was started,
for their hospitality. This work was supported by a Research Fellowship
from St.John's College, Cambridge.

\end{document}